\documentclass[conference]{IEEEtran}
\IEEEoverridecommandlockouts

\pdfoutput=1

\usepackage{cite}
\usepackage{amsmath,amssymb,amsfonts}
\usepackage{algorithmic}
\usepackage{graphicx}
\usepackage{textcomp}
\usepackage{xcolor}
\usepackage{authblk}
\usepackage[nolist]{acronym}
\usepackage[utf8]{inputenc}
\usepackage{url}
\usepackage{breakurl}
\usepackage[breaklinks]{hyperref}

\def\BibTeX{{\rm B\kern-.05em{\sc i\kern-.025em b}\kern-.08em
    T\kern-.1667em\lower.7ex\hbox{E}\kern-.125emX}}
    
\begin{document}

\title{From Tether to Libra: Stablecoins, \\ Digital Currency and the Future of Money}

\author[1]{Alexander Lipton}
\author[2]{Aetienne Sardon}
\author[3]{Fabian Sch\"{a}r}
\author[2]{Christian Sch\"{u}pbach}
\affil[1]{Sila, The Hebrew University of Jerusalem, Massachusetts Institute of Technology}
\affil[2]{Swisscom, Digital Business Unit, Fintech}
\affil[3]{Center for Innovative Finance, University of Basel}
\affil[ ]{\textit {alex@silamoney.com, aetienne.sardon@swisscom.com, 
f.schaer@unibas.ch, christian.schuepbach@swisscom.com}}
\maketitle

\begin{acronym}[aligator]
    \acro{DLT}{Distributed Ledger Technology}
    \acro{ECB}{European Central Bank}
    \acro{BIS}{Bank for International Settlement}
    \acro{FINMA}{Swiss Financial Market Supervisory Authority}
    \acro{FINMA's}{Swiss Financial Market Supervisory Authority's}
    \acro{DApp}{Decentralized Application}
    \acro{DeFi}{Decentralized Finance}
    \acro{CDP}{Collateral Debt Position}
    \acro{EMI}{Electronic Money Institution}
    \acro{KYC}{Know Your Customer}
    \acro{AML}{Anti-Money-Laundering}
    \acro{AMLA}{Anti-Money Laundering Act}
    \acro{FMIA}{Financial Market Infrastructure Act}
    \acro{CISA}{Collective Investment Schemes Act}
    \acro{IMF}{International Monetary Fund}
    \acro{IMF's}{International Monetary Fund's}
    \acro{SME}{Small and Medium-Sized Enterprise}
\end{acronym}

\section{Introduction}
What first started as a niche phenomenon within the cryptocurrency community has now reached the realms of multinational conglomerates, policy makers, and central banks.\\

From JP Morgan's Jamie Dimon to Facebook's Mark Zuckerberg, stablecoins have made their way onto the agenda of today's top CEOs. As projects like Libra have enjoyed broad media coverage they are also increasingly scrutinized by regulatory authorities, \cite{EU}\cite{Congress}\cite{FINMA}. And as the term ``stablecoin'' spread, its meaning started to blur. This is problematic. An unclear definition may make us susceptible to deceptive innovation, that is, reintroducing existing services but in a different appearance. We ought to ask ourselves: are stablecoins here to stay or are they simply old wine in new bottles?\\

This article aims to educate on stablecoins by providing a historical context on their origin and describing  which key factors have been driving their adoption. Moreover, we review existing terminologies and taxonomies on stablecoins and examine their disruptive potential. Based on this, we propose a novel definition of stablecoins and outline an alternative taxonomy. We briefly discuss the different use cases of stablecoins, as well as the underlying economic incentives for creating them. We also touch upon regulatory considerations and briefly summarize key factors driving future development.

\section{Motivation}
Money is omnipresent in modern life, but we rarely dare to question it. As it has existed for more than 5,000 years, one is prone to misconceive it as a fixed concept, when, in fact, it has been continuously changing. And as our society evolves so does the way we interact and transact with money. With new forms of money on the rise, we are challenged to question our understanding of money and ask ourselves: how shall the future of money look like?\\

Stablecoins have been discussed as a potential candidate for a new, faster, more accessible and transparent form of money. With Facebook's engagement in the Libra stablecoin, there has been substantial attention to the topic. But the emergence of new technologies, such as \ac{DLT}, has subtly diverged our focus away from "how can we create value" to "how can we use this technology".\\

In order to prevent falling prey to deceptive innovation, policymakers, incumbents, challengers and the general public alike should have the interest to develop a sound understanding of stablecoins. 

\section{WIR: A Stablecoin Precursor}
The concept of devising a supplementary currency system is not a new one. One of the most successful examples is the Swiss WIR Bank, formerly known as the Swiss Economic Circle. WIR was founded in 1934 by Werner Zimmermann and Paul Enz, \cite{Orell}. WIR is the abbreviation for \textit{Wirtschaftsring-Genossenschaft} (meaning ``mutual economic support circle'') but at the same time also means ``we'' in German, \cite{Gimigliano}. WIR was driven out of the ambition to alleviate the negative effects of the Great Depression, solve the associated middle class crisis and reform the monetary system on the basis of the \textit{Freigeld} (meaning ``free money'') theory, \cite{Vuillaume}. To achieve these goals WIR initiated its own WIR currency (``CHW''), allowing participants to exchange goods and services without using conventional fiat currencies.

Today the WIR network comprises over 62,000 \acp{SME}, reporting a turnover of almost 2 billion WIR in 2012. While nowadays the WIR network is also open to private clients, its focus remains on \acp{SME}. The main benefits of joining the WIR network are threefold: first, companies joining on average experience a 5\% increase in business, most likely due to loyalty effects, \cite{Orell}. Secondly, participants in the WIR network can obtain loans at a lower interest rate than compared to traditional bank loans. And thirdly, members of the WIR network experience a greater sense of solidarity and community, \cite{Vuillaume}.

Companies participating in the WIR system commit to accepting CHW for their goods and services at a 1:1 rate to the corresponding CHF amount. In order to join the WIR system, companies can apply for a zero interest loan of up to 10,000 CHW, \cite{WIR2}. If a company wishes to leave the system, surplus CHW must be spent within the system. While the WIR system bears  similarities with the idea of a stablecoin, there also is a notable difference: buying or selling CHW on a secondary market is strictly prohibited, \cite{WIR}. In section \ref{sec:definition} we will come back to identify the presence of a secondary market as one of the key characteristics of a stablecoin. 

\section{A Brief History of Stablecoins}
\label{sec:history}
It is impossible to have a well-rounded discussion on stablecoins without examining their origins. Although numerous stablecoin projects exist today, there is one stablecoin that stands out in its significance, i.e., Tether, \cite{Consensys}. As one of the first and to this day most widely used stablecoins, Tether has played a significant albeit controversial role for the development of stablecoins.\\

\begin{figure}[htbp]
	\centerline{
		\includegraphics[width=\linewidth]{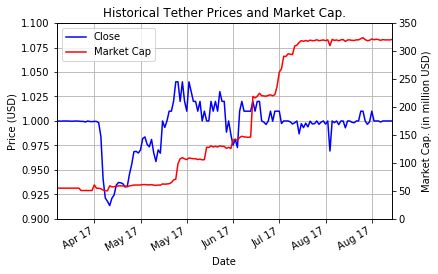}
		}
	\caption{Tether Prices and Market Cap.}
	\label{fig:tether}
\end{figure}

As of December 2019, there are more than 4.1 billion Tether tokens in circulation. Each Token is supposed to be worth \$1. The issuing company, Tether Limited, claims that Tether tokens are 100\% backed by liquid Reserves. However, numerous parties raised allegations that there is a shortfall in its reserves. These allegations have been fuelled by severe deficiencies in
the auditing process, \cite{Tether}\cite{Forbes}. Doubts about Tether's reserves have repeatedly manifested themselves in lower secondary market prices. For example, at the beginning of 2017 Tether's secondary market price dropped to as low as \$0.91 (see figure \ref{fig:tether}). Nonetheless, Tether is still by far the most actively traded stablecoin. In fact, in terms of trading volume, it can easily compete with other cryptoassets such as Bitcoin or Ether.\cite{Coinmarketcap}\\ 

Tether, known initially as Realcoin, was founded in 2014 by Brock Pierce, Craig Sellars and Reeve Collins. Before founding Tether, two of the three co-founders worked on a project called Mastercoin (later rebranded as Omni). Mastercoin's mission was to allow users to create their own virtual currencies on top of the Bitcoin protocol, \cite{Zynis}\cite{Sellars}, \cite{Roth}. For this purpose, the Mastercoin Foundation developed an additional layer on top of Bitcoin, which would later serve as the technological foundation for issuing the first Tether tokens in October 2014.\\

One of the key drivers for Tether's growth was its listing on and distribution through cryptocurrency exchanges. Bitfinex, as one of the largest cryptocurrency exchanges, played a pivotal role in promoting Tether. Although denied by both companies' officials, multiple indications suggest Bitfinex and Tether have been closely affiliated.\footnote{For example, the Paradise Papers leak revealed that Bitfinex officials were responsible for setting up Tether Holdings Limited in the British Virgin Islands in 2014. Moreover, Tether Limited and Bitfinex both share the same CEO, CFO and general counsel, \cite{Tether2}\cite{Bitfinex}.}\\

When Tether tokens first started trading on Bitfinex in 2015, their turnover was rather insignificant. However, as cryptocurrencies gained traction, so did the Tether stablecoin. By mid-2017 Tether's market capitalization had surpassed \$100mn (see figure \ref{fig:tether}). At the same time Bitfinex users were starting to experience substantial delays in their U.S. Dollar withdrawal requests, \cite{Bloomberg.} Shortly thereafter rumors spread that Bitfinex had been cut off from its U.S. Dollar wire transfers.\footnote{As it later turned out Bitfinex's bank accounts had been frozen by Wells Fargo, \cite{Buntinx}.} At the same time numerous cryptocurrency exchanges, such as Kraken, Binance and Huobi, decided to list Tether trading pairs, \cite{Kraken}\cite{Binance}\cite{Huobi}. This support allowed Tether tokens to spread across the cryptocurrency trading ecosystem quickly. Tether allowed to circumvent traditional wire transfers by providing an alternative settlement mechanism. Although, token users were unable to withdraw their U.S. Dollars, Tether allowed them to transfer their U.S. Dollar-pegged tokens between exchanges, without being exposed to the price volatility of cryptocurrencies.\\

After the 2018 cryptocurrency crash, a paper was published claiming that Tether was used to inflate and manipulate Bitcoin prices, \cite{Griffin}. It has been suggested that cryptocurrency exchanges may have had a vested interest to continue the distribution of Tether and in general promote the use of stablecoins to increase trading volumes. Moreover, stablecoins posed an opportunity for cryptocurrency exchanges to become less dependent on unstable banking relationships, \cite{Upson}.\\

Given the strong demand for a stablecoin like Tether, it comes as no surprise that new players rushed into the market from late 2017 onwards. For example, in 2018, TrustToken, Paxos, Gemini, and Circle all launched a U.S. Dollar pegged stablecoin. These projects promoted their stablecoins as being more reliable and trustworthy, providing higher transparency in terms of their reserve management, \cite{Gemini}\cite{Coinbase}\cite{Huobi2}. Note that all of these stablecoins were primarily designed to strive within the cryptocurrency space. The surge in projects also sparked creativity in terms of how to design a stabilization mechanism for a stablecoin. For example, a project called Maker DAO built a decentralized stablecoin (DAI) whose reserve would be comprised of other cryptocurrencies and completely governed on-chain through Ethereum smart contracts. Another project called Basis raised \$133 million with the goal of launching an algorithmic cryptocurrency protocol that claimed to create a stable digital currency without requiring any asset backing whatsoever. However, it is noteworthy that the Basis team decided to shut down the project because it would have been applicable to U.S. securities regulation, \cite{Basis}.\\

In parallel to the stablecoin developments from the cryptocurrency community, large cooperations started to experiment with blockchain technology -- mainly for large scale transactions. For example, UBS published a paper introducing the so-called Utility Settlement Coin in 2016, which financial institutions can use for facilitating cross-border payments and settlement, \cite{UBS}. In 2018, the MIT developed the idea of Tradecoin, in which multiple ``sponsors'' form a consortium where they can tokenize their assets and build a system of digital cash on top of that. The sponsors contribute assets to a collectively owned asset pool and receive Tradecoins in exchange from the consortium. The safekeeping of the consortium's asset pool is managed by a narrow bank to guarantee the full-backing of the Tradecoins with the actual asset base. The consortium can then use their Tradecoins as an asset base to issue e-cash tokens to retail users, \cite{Lipton}, \cite{Lipton3}, \cite{Lipton4}. At the beginning of 2019 JP Morgan announced that it would become the first U.S. bank to create a digital coin representing a fiat currency. While these projects are not necessarily comparable to a stablecoin like Tether, they do appear to have been fuelled by the associated rising interest for novel digital currency forms. 

In June 2019, Facebook officially revealed its plans to launch a new global digital currency called Libra, \cite{WSJ}. The Libra project immediately triggered strong headwinds from regulators. For example, France's finance minister Bruno Le Maire said that ``no private entity can claim monetary power, which is inherent to the sovereignty of nations.'' \cite{Reuters}. Publications from the \ac{ECB} and \ac{BIS} followed shortly in August and October 2019, discussing potential risks associated with stablecoins.

\section{Terminology}
In this section, we first briefly discuss the etymology of stablecoins and then review the strengths and weaknesses of standard stablecoin definitions. We then point out some of the difficulties surrounding stablecoin terminology. We continue by briefly reviewing Christensen's Theory of Disruptive Innovation in the context of stablecoins and finally provide a new definition of stablecoins.

\subsection{Etymology - From Bitcoin to Stablecoin}
Before 2008 the term "coin" was unambiguously associated with actual physical coins. The advent of Bitcoin, however, lead to a recontextualisation of the word. One can only wonder why Bitcoin was not named Bitcash or Bitmoney at the time.\footnote{Perhaps archaic words were more aligned with the ideological tendencies of declinism and crypto-anarchism within the cryptocurrency community.}\footnote{Note that both Bitcash and Bitmoney exist today but were launched long after Bitcoin in early 2019 and late 2018.} But as Bitcoin emerged, the word coin experienced a semantic change. Its usage was now broadened to the digital economy.\\

As the number of cryptocurrency projects increased, so did the excitement for the digital coin jargon. From Litecoin to Dogecoin, digital coin minting proved very popular. With a plethora of inherently volatile digital coins the blockchain community started exploring whether blockchain could also be used to create more stable cryptocurrencies, or, in other words \emph{stable coins}.\\ 

\begin{figure}[htbp]
	\centerline{
		\includegraphics[width=\linewidth]{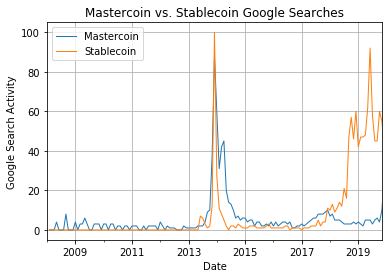}
		}
	\caption{Google searches going from "Mastercoin" to "Stablecoin".}
	\label{fig:mastercoin}
\end{figure}

Data from Google Trends shows that the term stablecoin first emerged in late 2013. Its appearance coincided with a spike in searches for Mastercoin (see figure \ref{fig:mastercoin}). As described in section \ref{sec:history}, Mastercoin laid the groundwork for Tether and made the until then vague concept of a stablecoin a reality. Thus, Mastercoin and stablecoins are closely intertwined, both from a conceptual but also from an etymological viewpoint.

\subsection{Introduction to Stablecoin Terminology}
While many different definitions of stablecoins exist, we highlight the one given by the \ac{ECB}:

\begin{quote}
	\textquotedblleft [Stablecoins are defined as] digital units of value that are not a form of any specific currency (or basket thereof) but rely on a set of stabilisation tools which are supposed to minimize fluctuations of their price in such currency(ies)\textquotedblright, \cite{ECB}.
\end{quote}

Although a rather broad definition, it reflects three important aspects:
\begin{enumerate}
	\item First, it is technology-neutral, and it excludes already existing distinct forms of currencies that simply use \ac{DLT} for recording purposes. This fact helps to differentiate between stablecoins as a genuinely new form of money (e.g., DAI) and commercial bank money that is powered by new technology (e.g., JPM Coin).
	\item Second, it highlights that there must be some form of stabilization mechanism to reduce volatility relative to an existing currency.
	\item And third, it points out that stablecoins have a market price of their own, implying that its price expressed in the target quote currency is not necessarily equal to one.
\end{enumerate}

Other definitions are often phrased in a way that blurs the lines between the stablecoin and the ``linked'' asset. For example, the \ac{BIS} states that ``stablecoins have many of the features of cryptoassets but seek to stabilize the price of the ``coin'' by linking its value to that of a pool of assets.'' The word ``link'' suggests a form of equivalence between the stablecoin and the ``linked'' asset, when in fact both need to be conceived as separate assets and can potentially be decoupled. In this respect, a stablecoins is to its ``linked'' asset as a derivative is to its underlying. In particular, most stablecoins introduce some counterparty risk.

\subsection{Motivation - Why new Terminology?}
\label{sec:disruption}
As already pointed out, there is a blurring line between stablecoins that are a genuinely new type of asset and those that represent existing forms of currency. We advocate to avoid introducing new terminology for already well-understood and existing concepts (e.g. commercial bank money). Instead, we endorse to use the term stablecoin to label and identify genuinely innovative forms of money that lay outside of the established monetary system (potentially beyond the control of central banks) but have the potential to fundamentally change and disrupt it.\footnote{For a more detailed comparison between the various already existing forms of money and genuine cryptocurrencies, we refer the reader to the monetary control structure cube proposed by \cite{Berentsen}.}\\

Christensen's Theory of Disruptive Innovation provides a useful tool to help identify potentially disruptive stablecoins and distinguish them from rebranded traditional financial services. According to Christensen's theory there are two forms of innovation: sustaining innovation and disruptive innovation.\cite{Christensen}\\

\begin{figure}[htbp]
	\centerline{
		\includegraphics[width=\linewidth]{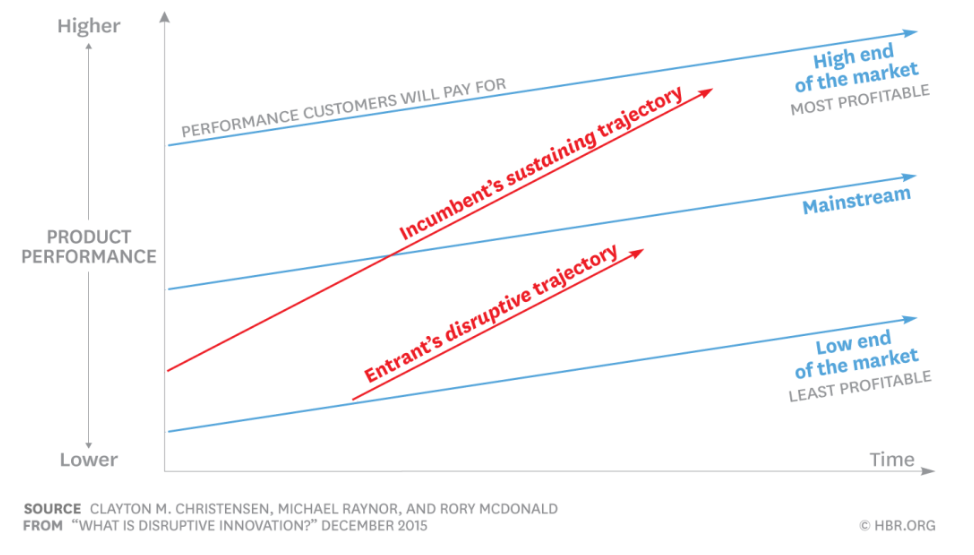}
		}
	\caption{Disruptive vs. sustaining innovation trajectories, \cite{Christensen}.}
	\label{fig:disruption}
\end{figure}

\begin{itemize}
	\item Sustaining innovation is aimed at improving existing products for an incumbent's established customer base. Typically, higher-quality products are introduced to satisfy the high-end of the market, where profitability is highest.
	\item Disruptive innovation, on the other hand, is initially considered inferior by most of an incumbent’s customers. It either starts in (1) low-end or (2) new-market footholds. (1) In the first case, a disruptor introduces a good-enough product for otherwise underserved low-end customers. (2) In the second case, a disruptor introduces a genuinely new product in a market where none existed, basically turning non-consumers into consumers. As shown in figure \ref{fig:disruption}, the disruptor then moves upmarket, providing the quality that mainstream customers require, while preserving the advantages that drove his early success. When mainstream customers start adopting the new product, disruption has occurred.
\end{itemize}

Putting Christensen's Theory of Disruptive Innovation into practice, let us consider three examples: Tether, JP Morgan Coin, TradeCoin, and Libra.

\begin{enumerate}
	\item Tether exhibits characteristics of a disruptive innovation. The reasons are as follows: first, it originated in a low-end market that was otherwise neglected by incumbents (see section \ref{sec:history}). Tether provided a good-enough product to help cryptocurrency users transact in something that is close-enough to the U.S. Dollar, without requiring access to traditional payment systems or banking services. Secondly, Tether started moving upmarket. Being listed on over 100 exchanges, including conservative ones such as Coinbase, both mainstream as well as high-end institutional customers (e.g., proprietary trading firms) have started using Tether. Besides, Tether is scaling up to support for additional blockchain networks (e.g., Ethereum, Liquid, Tron) and currencies (e.g., EUR, CNY).
	\item JPM Coin displays the qualities of a sustaining innovation. There are two reasons to support this view: first, the coin is aimed at making inter-bank clearing and settlement better. Such services have been available before JPM Coin, but the coin was introduced to make the process faster and more efficient. Secondly, the target customer base is clearly in the high end of the market (as JPM Coin is exclusively available to institutional clients) and not geared towards mainstream or low-end customers. Therefore, JPM Coin follows an incumbent's sustaining innovation trajectory (see figure \ref{fig:disruption}).
	\item TradeCoin represents a disruptive innovation. Its main objective is to give asset-backed currencies a new lease of life. In its mature state, the DTC can serve as a much-needed counterpoint to fiat reserve currencies of today.
	\item Libra's innovation quality depends on its go-to-market strategy. It may be considered a disruptive innovation if (1) it indeed initially focusses on the low-end market and (2) subsequently moves up-market. According to its website, Libra's vision is to provide payment services for the 1.7 billion unbanked. The unbanked population is a low-end market that traditionally has been neglected by incumbents. With each of Libra's founding members having a global reach and substantial financial resources, they are best equipped to turn their vision into a reality. If Libra delivers on its promise and someday dominates the unbanked market, it has a strong potential to move upmarket and eventually disrupt traditional payment services.
\end{enumerate}

While missing out on a potential sustaining innovation may only have minor repercussions, failing to detect a disruptive innovation poses an existential threat to an incumbent's business. With the rise of \ac{DLT}, the financial services space has been overcrowded with seemingly innovative payment solutions. At the same time, it has become increasingly difficult to separate genuinely new payment solutions from re-engineered legacy systems under the guise of innovation. We, therefore, advocate to use the term stablecoin more carefully to label genuinely new forms of money that possess disruptive potential. At the same time, we suggest avoiding the term stablecoin to relabel existing products or minorly improved ones.

\subsection{A Novel Definition}
\label{sec:definition}
We aim to provide a compact definition for stablecoins that captures their essential characteristics and is easy to use. We identify three fundamental properties that characterize a stablecoin and sets it apart from other forms of money.\\

A stablecoin is a digital unit of value with the following three properties:
\begin{enumerate}
	\item It is not a form of currency, 
	\item it can be used without any direct interaction with the issuer,
	\item it is tradable on a secondary market and has a low price volatility in terms of a target quote currency.
\end{enumerate}

The advantages of using this definition are as follows: first, it is technology-neutral, focusing on the underlying conceptual elements of a stablecoin instead of its implementation details. Secondly, it is mutually exclusive to existing forms of currency (similar to the \ac{ECB} definition). This property makes the definition useful in identifying genuinely new forms of money with disruptive potential. Thirdly, it highlights the unique features that make stablecoins distinct from previously known payment systems. Stablecoins can be used without requiring any direct interaction with the issuer (e.g., for peer-to-peer transfers) and they can be exchanged on a secondary market at a somewhat reliable and ``stable'' price.\footnote{In fact, some stablecoins require an efficient secondary market for their stabilization mechanism to work. Such stablecoin systems are constructed in a way that, by-design, arbitrage opportunities arise as soon as the market price deviates from the target par value.}

\section{Taxonomy}
\label{sec:taxonomy}
\subsection{Review}
Most taxonomies classify stablecoins based on differences in their collateralization-mechanics. For example, some authors suggest to differentiate between fiat-, commodity-, crypto- and un-collateralized stablecoins, \cite{Hays}. Others suggest to group by on-chain-, off-chain- and un-collateralized stablecoins, \cite{Berentsen2}. Again others distinguish between fully fiat collateralized, partially fiat collateralized, crypto overcollateralized, dynamically stabilized and asset collateralized stablecoins, \cite{Lipton2}. Since taxonomies focused on collateralization-types are already well known, we will refrain from repeating them. Instead of focusing on collateralization setups, we point to a simple yet revealing dichotomy of stablecoins: i.e., does the stablecoin represent a legal claim and therefore require a functioning legal system or does it work, even in the absence of any institutions. The former are issued as an IOU and the issuers may be held responsible if they fail to deliver on their promise. The later are self-sustained in the sense that the stabilization mechanism does not rely on any institutions nor a functioning legal system. Frequently examined features such as the degree of decentralization and openness of a stablecoin system are highly correlated with the existence or absence of a coin holders' legal claim. For example, if there is no legal claim associated with a coin, the system is most likely to be decentralized, with low accountability and high openness. As a stablecoin system strives through its network effects, it is unlikely to be restricted for its own sake, but because of regulatory and legal constraints. If, on the other hand the legal and regulatory structure allows the system to be open, it most likely will be.

\subsection{A Tripartite Classification}
The fair value of a stablecoin should be equal to its expected redeemable amount. The trust in the redeemability of a stablecoin may be based on different rationales. As an expansion of existing taxonomies, we provide an additional classification to reflect these differing redeemability rationales, grouping stablecoins into three categories:

\begin{enumerate}
	\item Claim based: these coins can come in two forms: first, coin holders can have a direct legally enforceable right to personally redeem their coins against a pre-defined amount or value of a reference asset (e.g., fiat money or commodity). For example, Circle states in its terms for USDC that ``sending USDC to another address automatically transfers and assigns to that Holder, and any subsequent Holder, the right to redeem USDC for U.S. Dollar funds'', \cite{Circle}. Moreover, coins that are structured as electronic money or commercial bank money in prepaid payment systems would also fall under this category. Secondly, coin holders may benefit from a transitive claim, i.e., they may not be entitled to redeem the coin themselves but instead have to go through a third party. For example, two-tiered stablecoin systems, such as proposed by Libra where some privileged users ("authorized resellers") have a right to redeem while other users do not, are based on the idea of a transitive claim.
	\item Good-Faith based:  these are coins where the holder believes in the good business practices of the issuer, assuming redeemability of his coins without having any legal right. The issuer typically promotes the coins to be backed by reserves but excludes any right of redemption in its terms and conditions. For example, TrustToken states in its legal terms that "the Company itself does not guarantee any right of redemption or exchange of TrueCurrency tokens for fiat currency, "\cite{TrustToken}.\footnote{Similarly, stablecoins that promote buybacks to stabilize market prices would fall under this category as well as such buybacks would typically occur at the issuer's discretion without being legally bound to it.}\item Technology based: these are coins where technology is used to autonomously induce price stabilization, e.g., using smart contracts to store and manage cryptoasset collateral. These systems do not rely on a legal claim or a user's faith in the good intentions of the issuing entity. Instead, the user's expectations of the redeemability is driven by their trust in the underlying technology and implementation. For example, on-chain collateralized and algorithmic stablecoins belong to this category.
\end{enumerate}

\begin{figure}[htbp]
	\centerline{
		\includegraphics[width=\linewidth]{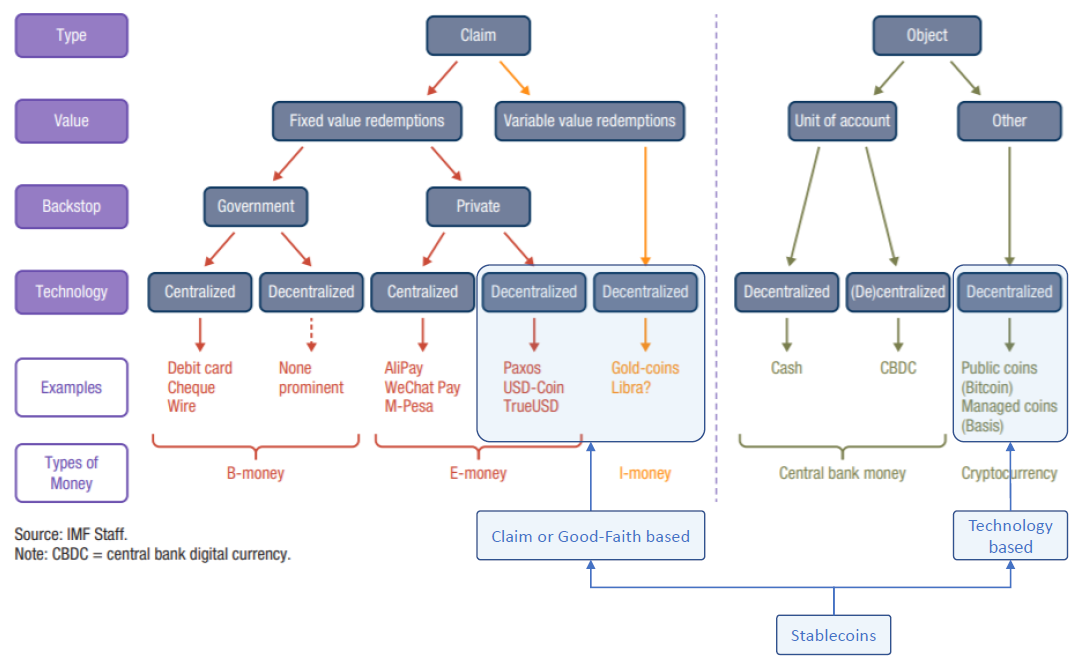}
		}
	\caption{Placing stablecoins into the IMF's ``money tree'', \cite{IMF}.}
	\label{fig:monetarytree}
\end{figure}

To put our view into a broader context, we refer to the \ac{IMF's} money tree (see figure \ref{fig:monetarytree}). The money tree identifies five different means of payment, i.e. B-money, E-money, I-money, Central bank money and cryptocurrencies. According to our definition (see section \ref{sec:definition}) and taxonomy, claim and good-faith based stablecoins would comprise I-money and partly E-money. Technology based stablecoins are congruent to the \ac{IMF's} definition of "managed coins", \cite{IMF}. We can see that our framework is at least partially congruent with the \ac{IMF's} categorization -- in particular with respect to differentiation based on the existence or absence of a claim. 

\section{Use Cases}
\label{sec:usecases}
Of the many use cases that have been discussed for stablecoins, the following have materialized so far:

\begin{itemize}
	\item Cross-border payments \& arbitrage: stablecoins have been used for cross-border payments, especially between cryptocurrency exchanges, giving traders a tool to take advantage of arbitrage opportunities and thereby improving market efficiency.
	\item Trading \& settlement: stablecoins have been used as a trading instrument to quickly convert from volatile cryptocurrencies into more stable currency substitutes and vice versa. Conversely, from the perspective of the cryptocurrency exchanges, they allowed them to offer their users U.S. Dollar-like trading and settlement functionalities without depending on traditional wire transfers. Thereby stablecoins enabled exchanges to become less reliant on often fragile banking partnerships. It is noteworthy that most stablecoin wallets are controlled by cryptocurrency exchanges, suggesting that users mainly transfer stablecoins between exchange omnibus wallets and rarely withdraw. For example, a recent report found that only about 300 entities control over 80\% of Tether tokens, with many of these being cryptocurrency exchanges, \cite{Kharif}.
	\item \ac{DeFi} applications: offer a broad variety of use cases including decentralized exchanges, lending markets, derivatives and on-chain asset management, \cite{schaer:20}. For all of these applications, stablecoins play an important role. Additionally, stablecoins like DAI allow users to take on leveraged trading positions.\footnote{For example, DAI currently uses an over-collateralization rate of 150\%, meaning that for \$100 worth of DAI \$150 worth of cryptoassets need to be locked in a so-called \ac{CDP} (or vault).\cite{Maker} A trader wanting to leverage his ETH position could lock \$100 worth of ETH and receive \$100/150 = \$66.6 worth of DAI. He could then buy \$66.6 worth of ETH, lock these assets in the \ac{CDP} and repeat this process again and again, yielding a total leveraged position of \$300.} Moreover, users can also lock up their DAI tokens to earn interest (e.g., Aave, Compound, dYdX).
\end{itemize}

Other use cases like payment, payroll, and remittance have not found much attention so far. Similarly, integration of stablecoins into \acp{DApp} or as a cash leg handling in smart contract-based financial contracts is yet to find wider adoption.

\section{Revenue Streams \& Cost Structure}
\label{sec:businesscase}
Stablecoin issuers may profit from multiple revenue streams. The composition of revenues may vary greatly depending on the exact stablecoin setup. For example, technology and claim-based stablecoins are likely to exhibit very different revenue stream structures. Regardless of the different stablecoin types and their differing revenue focus, we identify the following five revenue streams:

\begin{itemize}
	\item Interest earnings: stablecoin issuers typically allocate all interest earnings generated from the reserve fund to themselves. Issuers are not required and, in some instances, even prohibited to pass on interest earnings to coin holders. For example, \acp{EMI}, such as Circle, are prohibited to grant interest. Depending on the legal structure of the stablecoin, issuers may have varying degrees of freedom in terms of the reserve fund management. Generally, issuers have an incentive to issue stablecoins for currencies that offer positive interest rates. For example, TrustToken supports stablecoins for USD, GBP, AUD, CAD, and HKD, all of which used to provide positive interest rates, ranging between 0.75\% and 3.95\% p.a., \cite{TrustToken2}. However, most of these rates have been cut to almost zero lately. Depending on the size of the reserve funds, interest earnings may be substantial. For example, let us assume Tether has \$4.1 billion in reserves. "The composition of the Reserves used to back Tether Tokens is within the sole control and at the sole and absolute discretion of Tether", \cite{Tether3}. Let us further assume Tether's liquidity management allocates 80\% of the reserves into U.S. Dollar money market funds with an annual percentage rate of 1.7\%. The float would generate earnings of \$55.8 million per year, \cite{SwissFundData}.
	\item Transaction fees: stablecoin issuers may charge fees for every transfer. For example, Tether's smart contract has a feature that allows to charge up to 20bps of the transaction value with a maximum fee of \$50 per transaction, \cite{Etherscan}. Let us assume the daily average number of transactions to be 100,000, with an average size of \$5,000, \cite{Etherscan2}. If Tether were to charge a 1bps transaction fee, this would result in revenues of \$18.2 million per year. However, Tether has not charged any transaction fees so far, because doing so would disincentivize using Tether coins, potentially leading to a shrinking reserve fund and diminishing interest earnings. Moreover, the transaction fee would only apply to on-chain transactions (excluding any intra-exchange transaction). So far, interest earnings seem to have outweighed potential earnings from transaction fees. As commented in their code, Tether most likely sees transaction fees as a means of last resort ("[…] if transaction fees ever became necessary", \cite{Etherscan3}).
	\item Issuance \& redemption fees: stablecoin issuers may charge fees for the issuance (minting) and redemption (burning) of stablecoins. For example, Tether charges 0.1\% per deposit and the greater of 0.1\% and \$1,000 per withdrawal, \cite{Tether4}. Obviously the issuer may use fees to steer the in- and out-flows to its stablecoin. This may become necessary if the issuer has limitations in terms of the reserve fund size or balance sheet (e.g. considering capital requirements for \acp{EMI} or restrictions given by an issuer's banking partner). Similarly, in case of liquidity shortages an issuer may discourage outflows by imposing higher withdrawal fees.
	\item Cross-selling: stablecoin issuers may cross-sell additional services that build upon their stablecoin. For example, some cryptocurrency exchanges are closely affiliated with stablecoin issuers (e.g., Bitfinex and Tether as described in section \ref{sec:history}). Stablecoins may serve as a mean to attract and facilitate trading on their platforms. Moreover, exchanges may also market-make their own stablecoins, providing additional revenue potential.
	\item Secondary tokens: technology-based stablecoins systems are often designed as a two-fold token model, where one token serves as the stablecoin, and the second provides some special functionality to interact with the stablecoin system. The second token is typically designed to increase in value with stablecoin usage. The initiators of the stablecoin system regularly allocate a proportion of these tokens to themselves to benefit from such value appreciation. For example, DAI has a special governance token (MKR) which is also needed to close \acp{CDP} (see section \ref{sec:usecases}). Whenever a user wants to regain access to his locked cryptoassets, he needs to pay interest in the form of MKR tokens, which subsequently get burned. As the supply of MKR tokens decreases over time, there will be a lower supply that may c.p. lead to higher prices.
\end{itemize}

As with the revenue streams, a stablecoin's cost structure differs across stablecoin types. In particular, the cost structure will heavily depend on whether the issuing entity is regulated or not. In general, we identify the following seven cost components:

\begin{itemize}
	\item Legal, regulatory \& compliance: various legal and regulatory clarifications may become necessary. Depending on the regulatory status of the issuer, e.g. \acp{EMI}, licensing costs may incur. Licenses may be necessary for every jurisdiction in which the stablecoin shall be made available. Moreover, compliance efforts, e.g. to ensure adherence with \ac{KYC} regulations and applicable \ac{AML} requirements, need to be considered.
	\item IT development: in case a public \ac{DLT} is used, the stablecoin issuer benefits from the openness and interoperability of the underlying \ac{DLT}. Development costs would mostly comprise setting up the smart contract. In contrast, integration into third-party systems such as wallets or exchanges does not involve additional efforts (e.g., by adhering to standards such as ERC20). 
	\item IT audits: when a stablecoin is based on a public \ac{DLT} the proper functioning of the corresponding smart contract is of critical importance. Typically, a stablecoin issuer mandates security experts to audit the smart contracts in order to assure that the contracts do not have any security flaws and work as expected.
	\item Financial audits: depending on the regulatory status of the stablecoin issuer, audits of its financial statements may be mandatory. Some issuers may voluntarily conduct financial audits to assure users that the reserves are managed responsibly.
	\item Banking services: depending on the nature of the stablecoin, the issuer may rely on banking services to store its reserves.
	\item Key management: the issuance and redemption of stablecoins involve some form of approval workflow. Especially for stablecoins that use a public \ac{DLT}, secure management of potential admin/issuer keys is of utmost importance.
	\item Insurance: in case a stablecoin is backed by physical assets, such as gold or bank notes, the corresponding storage would typically require insurance coverage.
\end{itemize}

Depending on the stablecoin category, the cost structure may tend to involve higher fix costs than variable costs, providing an attractive scalable business case. It comes at no surprise that some stablecoin issuers are incorporated in offshore locations to evade regulatory requirements while still benefiting from the global scale that borderless \ac{DLT} systems provide. 

\section{Regulatory Considerations}
\label{sec:regulation}
From a regulatory standpoint, no unified definition of stablecoins exists so far. In order to reflect the current situation, we briefly review the \ac{FINMA} statement on stablecoins, the U.S. proposed "Managed Stablecoins are Securities Act of 2019" and the \ac{ECB} position.\\

\ac{FINMA} points out that no specific regulation currently covers stablecoins. However, following a technology-neutral approach, \ac{FINMA} states that many proposed stablecoin projects give rise to licensing requirements under the Banking Act or the \ac{CISA}. Moreover, as stablecoins are regularly intended to serve as a means of payment, the \ac{AMLA} is almost always applicable, resulting in strict \ac{KYC} requirements, transaction monitoring etc. Lastly, if a payment system of significant importance is to be created, a licensing requirement under the \ac{FMIA} is probable. \ac{FINMA} identifies eight categories of stablecoins with most of them falling under existing regulations. For example, if a stablecoin is linked to a fiat currency, this likely constitutes a deposit-taking business under banking law (e.g., Tether). If a stablecoin is linked to a basket of fiat currencies, as proposed by Libra, the applicable regulation depends on who bears the market risk associated with the management of the currency basket. If the issuer bears it, this  constitutes a deposit-taking business under banking law. If the token holders bears it, the stablecoin is considered a collective investment scheme, \cite{FINMA2}. The fact that \ac{FINMA} subsumes most of the stablecoins under existing regulation substantiates our view that in many instances stablecoins are not a new form of currency (see sections \ref{sec:definition}, \ref{sec:disruption} and \ref{sec:taxonomy}). Similar to \ac{FINMA's} substance over form attitude, U.S. policymakers have advocated a "same risks, same rules" approach towards stablecoins. In the newly proposed "Managed Stablecoins are Securities Act of 2019", they define "managed stablecoins" as a digital asset whose value is determined by reference to the value of a basket of assets and where the holder is entitled to obtain payment based on the value of that basket. These "managed stablecoins" shall be considered as securities under the existing Securities Act of 1933, \cite{Congress}.\\

Lastly, the \ac{ECB} has formulated a similar stance towards the regulatory treatment of stablecoins as \ac{FINMA} and U.S. policymakers. While the \ac{ECB} does acknowledge that some stablecoins may fall outside of current regulatory regimes, in many cases the risks that they entail are ``the same as for their non-\ac{DLT} competitors.'' In particular, the \ac{ECB} states that stablecoins issued as tokenized funds are likely to qualify as electronic money and as such are already covered by the existing second Electronic Money Directive (EMD2) in the EU. The \ac{ECB} also points out that the use of a new technology may often be mistaken for the introduction of a new asset class. However, those stablecoins that are truly part of the new phenomenon of crypto-assets, may still involve major uncertainties relating to their governance and regulatory treatment, \cite{ECB}.\\

It is noteworthy that the regulatory treatment of stablecoins is not only relevant for potential licensing requirements but may also have an impact on its accounting treatment. For example, concerning the asset side of a bank's balance sheet, \ac{FINMA} has suggested applying an 800\% risk weight to cryptocurrency assets regardless of whether these assets are held in the bank or trading book, \cite{Mauchle}. Depending on its specific nature, a stablecoin may be seen as a cryptocurrency and thus induce higher capital requirements. Since no official statement has been given so far, it seems that every bank intending to transact in stablecoins is best advised to check the coin's regulatory qualification with the regulator. Similarly, from the perspective of a stablecoin issuer, the accounting treatment on the liability side of his balance sheet may vary greatly depending on the exact nature of the stablecoin. While stablecoins construed as electronic money may be straightforward to account for, more exotic stablecoin formats may be rather challenging.

\section{Possible Scenarios \& Outlook}
\label{sec:outlook}
Factors from within and outside of the stablecoin universe are going to drive further development of stablecoins. Incumbents and policymakers, as well as challengers and users, are going to influence the stablecoin evolution.\\

Within the cryptocurrency universe, the creation of a new prominent \ac{DApp} or cryptoasset may lead to a sudden demand shock in stablecoins. For example, if stablecoins are required to interact with such a DApp or represent the only access point to purchasing a new promising cryptoasset, the demand for stablecoins would likely surge. Similarly, the adoption of decentralized exchanges may as well lead to increasing demand for stablecoins in order to facilitate trading. In contrast, in case of a major incident, such as the detection of a critical vulnerability in a \ac{DLT} system or a large-scale security breach with one of the dominant cryptocurrency exchanges, stablecoins are likely to backdrop. Depending on the severity of such an incident, policy makers may see themselves forced to impose stricter rules on businesses interacting with stablecoins. A policy shock, e.g., the introduction of specific licensing requirements, may make stablecoin projects less attractive, and in the worst case, bring further development to a halt. Aside from the regulatory circumstances, the overall economic environment may impact stablecoin adoption. For example, if interest rates were to normalize, the demand for more risky asset classes may recede and lead to higher opportunity costs for users when holding zero-interest bearing stablecoins. On the other hand, in case of a financial crisis, users may suddenly find themselves attracted to alternative forms of currency such as cryptocurrencies. Increased trading activity on cryptocurrency exchanges could positively affect the popularity of stablecoins. Lastly, the overall monetary system may be fundamentally changed through the introduction of a central bank digital currency (CBDC), potentially upstaging stablecoins.

\section{Summary \& Conclusion}
Stablecoins are an ambiguous concept of money. While they first originated in the world of cryptocurrencies, they have now become an independent concept of their own. Nonetheless, it is vital to understand how and why they first came into existence in order to develop a deeper understanding of what problems they currently solve and may address in the future.\\

As described in section \ref{sec:history}, stablecoins developed initially from the idea of democratizing the issuance of private currencies. At the same time, cryptocurrency exchanges needed a fiat currency substitute that would allow them to become less reliant on typically fragile banking partnerships. Stablecoins proved to be an elegant solution for growing the cryptocurrency trading ecosystem while minimizing dependency on traditional banking services. As Tether grew popular, so did the general enthusiasm for stablecoins. And as the usage of the term stablecoin spread, its meaning started to blur.\\

But an imprecise terminology may make us susceptible to deceptive innovation, overestimating the significance of re-engineered legacy payment systems and potentially overlooking more profound changes in our monetary system. As described in section \ref{sec:disruption}, Christensen's Theory of Disruptive Innovation provides a useful tool to distinguish between stablecoins as a genuinely new asset type and old wine in new bottles. Building upon those insights, we provide a new definition that distills the essential characteristics of a stablecoin. Specifically, we claim that: (1) it is not an existing form of currency, (2) it does not require any direct relationship with the issuer, and (3) it is tradable on a secondary market at a relatively stable and predictable price.\\

In section \ref{sec:taxonomy}, we propose an easy to use yet expressive taxonomy that focuses on the absence or existence of a legal claim, distinguishing between claim-, faith- and technology-based stablecoins. We also put this classification into a broader context by comparing it with existing taxonomies and find a strong congruence with the IMF's money tree.\\

In section \ref{sec:usecases}, we briefly review current use cases of stablecoins, highlighting cross-border payments, cross-cryptocurrency exchange settlement and \ac{DeFi} applications. We claim that the idea of stablecoins has outgrown its cryptocurrency origins. However, its usage is still very much rooted in the cryptocurrency space.\\

In section \ref{sec:businesscase}, we examine the revenue and cost structure of stablecoins. We find that interest earnings on the reserve funds provide substantial upside potential for stablecoin issuers. Given their predominantly fixed-cost structure, stablecoins constitute highly scalable business models. Unsurprisingly, to reduce costs, many issuers are incorporated in offshore locations while still benefitting from the global reach of today's \ac{DLT} platforms.\\

In section \ref{sec:regulation}, we briefly consider the regulators' perspectives by reviewing statements given by the \ac{FINMA}, U.S. policymakers as well as the \ac{ECB}. We find that most regulators have a technology-neutral view, aiming to subsume stablecoins under existing regulations.\\

In section \ref{sec:outlook} we complete our stablecoin examination by shortly outlining potential future scenarios. We find that claim- and faith-based stablecoins build upon existing money forms, whereas technology-based stablecoins are decoupled from the traditional money creation circle.\\

All in all, we conclude that stablecoins are a moving target with tremendous potential to fundamentally change the financial system. With \ac{DLT} providing a borderless and easy to integrate infrastructure, stablecoins have the potential to scale rapidly on a global scale and disrupt existing payment systems. Stablecoins are challenging our notion of money, creating a paradox situation in which they may be used like a currency without actually being labelled a currency. It remains yet to be seen whether stablecoins are going to coexist, complement, or takeover existing payments. But in any case, we should aim to use a more concise technology-neutral language, allowing us to focus on the truly disruptive potential of future money forms and applying new technologies such as \ac{DLT} in a more purpose-driven way. 

\bibliographystyle{IEEEtran} \bibliography{IEEEabrv,whitepaper} 

\end{document}